\documentclass{ws-procs975x65}
\usepackage{graphicx}
\newcommand{\Vh}{\hat{V}}
\newcommand{\Kh}{\hat{K}}
\newcommand{\finn}[1]{\phi^{\pm}_{#1}}

\newcommand{\eb}{\frac{e^{-|\alpha|^2} |\alpha|^{2 n}}{n!}}
\newcommand{\ebbb}{\frac{e^{-3|\alpha|^2} |\alpha|^{2 (l+n+m)}}{l!m!n!}}
\newcommand{\ass}{\alpha}
\newcommand{\as}{\alpha^*}
\newcommand{\fb}{\bar{f}}
\newcommand{\gb}{\bar{g}}

\newcommand{\sz}{\hat{s}_{z}}
\newcommand{\sy}{\hat{s}_y}
\newcommand{\sx}{\hat{s}_x}

\newcommand{\six}{\hat{\sigma}_x}
\newcommand{\siz}{\hat{\sigma}_{z}}
\newcommand{\siy}{\hat{\sigma}_y}
\newcommand{\vhsig}{\vec{\hat{\sigma}}}
\newcommand{\hsig}{\hat{\sigma}}
\newcommand{\hH}{\hat{H}}
\newcommand{\hU}{\hat{U}}

\newcommand{\hV}{\hat{V}}

\newcommand{\hK}{\hat{K}}

\newcommand{\hN}{\hat{N}}

\newcommand{\hro}{\hat{\rho}}
\newcommand{\vro}{\vec{\rho}}
\newcommand{\hR}{\hat{R}}
\newcommand{\half}{\frac{1}{2}}

\newcommand{\eps}{\varepsilon}

\newcommand{\BEQ}{\begin{equation}}
\newcommand{\EEQ}{\end{equation}}
\newcommand{\BEA}{\begin{eqnarray}}
\newcommand{\EEA}{\end{eqnarray}}
\newcommand{\sph}{spin-$\frac{1}{2}$ }
\newcommand{\ad}{\hat{a}^{\dagger}}
\newcommand{\add}{\hat{a}}
\newcommand{\spp}{\hat{\sigma}_+}
\newcommand{\smm}{\hat{\sigma}_-}
\newcommand{\fin}[1]{|\phi^{\pm}_{#1}\rangle}

\newcommand{\RI}{\hat{{\cal{R}}}_{0}}
\newcommand{\Rt}{\hat{{\cal{R}}}_{\tau}}

\renewcommand{\thesection}{\arabic{section}}
\renewcommand{\theequation}{\thesection\arabic{equation}}
\begin{document}
\title{SIMULTANEOUS MEASUREMENT OF NON-COMMUTING OBSERVABLES}
\author{G. AQUINO$^*$ and B. MEHMANI}
\address{Institute for Theoretical Physics, University of Amsterdam,\\
Valckenierstraat 65, 1018 XE,
Amsterdam, The Netherlands\\
$^*$E-mail: gaquino@science.uva.nl}
\begin{abstract}
It is shown that the full unknown
state of a spin-$\half$ system,  \textrm{S}, which, within Born's statistical interpretation,
is meant as the state of an ensemble of identically prepared systems  and is described by its density matrix, can be determined  with a simultaneous measurement
with the help of an ``assistant'' system \textrm{A}
whose initial state  is known. The idea is to let \textrm{S} and
\textrm{A} interact with each other in a known way during a proper
interaction time $\tau$,  to measure simultaneously two
observables, one of \textrm{S} and one of \textrm{A}  and  their correlation.
 One thus determines the three unknown components of
the polarization vector of \textrm{S} by means of repeated
experiments using a unique setting. In this way one can measure
simultaneously all the non-commutative observables of \textrm{S}, which might seem prohibited in quantum mechanics.
\end{abstract}

\bodymatter

\section{Introduction}\label{sec1}
The determination of the unknown state of a quantum system is one
of the most important issues in the field of quantum
information\cite{Hillery,Bennett}.
 This determination
involves a measurement process in which a macroscopic system,
apparatus, is coupled to the quantum system; during this process
the state of both the apparatus and the system is modified
\cite{measurement}. For instance, as currently described in many
textbooks, the $\textit{z}$-component of the polarization vector
of a spin-$\half$system, \textrm{S}, can statistically be
determined by means of a repeated Stern-Gerlach experiment. In
this process, the $\textit{x}$- and $\textit{y}$-component of the
polarization vector are destroyed as a consequence of the
non-commutation of the spin operator in the transversal directions.
Other  experimental settings seem therefore necessary to
measure the unknown polarization vector of \textrm{S}. Its three
components are represented by incompatible observables, the Pauli
operators, and their direct determination requires three
macroscopic apparatuses, differing by  a change of orientation
of the magnets and detectors. Likewise, the state of any two-level
system, represented by a $2 \times 2$ density matrix $\hro$ can be
fully determined only through measurement of three linearly
independent observables
which do not commute and cannot be simultaneously measured.\\
Nevertheless, we will prove that the whole unknown density matrix
of such a system \textrm{S}, in particular the full polarization
vector of a spin-$\half$ system, can be determined indirectly by
means of a set of measurements performed simultaneously on
\textrm{S} and an auxiliary system, \textrm{A}, which we term the
assistant. The strategy is the following: initially \textrm{S}
is in the unknown state that we wish to determine, while the
assistant \textrm{A} is in some known state. During some time
interval \textrm{S} and \textrm{A} interact in a known fashion. Their
joint state is modified, involves correlations and keeps memory
of the initial state of \textrm{S}. A simultaneous measurement of
one of the observables of \textrm{S} and \textrm{A} is then
performed. Repeating this process provides then three statistical data:
the expectation values of these observables and their correlation.
We will show that one can infer the three components of the
initial polarization vector of \textrm{S} from
the three data.\\
There are two approaches to this problem.  Either using another two-level assistant, for solid state applications \cite{PRL}  or using
an electromagnetic field as an assistant, for quantum optics applications.\\
After defining the general problem, we illustrate, in section 3,  the first approach, i.e. using another spin-$\half$ system as an assistant in a known pure state to determine the initial state of \textrm{S}.
Then, in section 4,  we show that it is also possible to use an electromagnetic field in a coherent state to determine the whole elements of the unknown
density matrix of \textrm{S}.

\section{Statement of the Problem}
The idea of mapping the state $\hro$ of an unknown spin-$\half$
system, \textrm{S}, onto a single observable of \textrm{S+A}
system by using another system in a known state $\hR$, \textrm{A},
was first proposed by D'Ariano \cite{D'Ariano}. It was explicitly
implemented in a dynamic form in Ref. ~\refcite{PRL}.

 The state of
the composite system $S+A$, which is tested, is \BEQ
\label{stateaftetau}\Rt = \hat{U} \RI \hat{U}^{\dagger},
 \EEQ
 where the initial state of $S+A$ is $\RI
= \hR \otimes \hro$ and the evolution operator is $\hat{U} = e^{ -
i \hH \tau}$. Therefore, the dynamics of the system yields the
required mixing of $\hro$ and $\hR$ and the simplest possible
non-degenerate observable of the composite system $\textrm{S+A}$,
$\hat{\Omega}$, can be chosen as a factorized quantity
\BEQ\label{factorizedObservables}
 \hat{\Omega} =
\hat{\omega}\otimes \hat{o}, \EEQ where the observable
$\hat{\omega}$ and $\hat{o}$ pertain to $\textrm{S}$ and
$\textrm{A}$ respectively. Then the spectral decomposition of
$\hat{\omega}$ and $\hat{o}$ can be used to construct the
projection operator $\hat{P}_{\alpha}$ of $\hat{\Omega}$ \BEA
\hat{\omega} =\sum_{i=1}^{m} \omega_i \hat{\pi}_i,
\;\;\;\;\;\;\;\;\hat{o} = \sum_{a=1}^{n} o_a\hat{p}_a, \EEA where
$\hat{p}_a$ and $\hat{\pi}_i$ are eigen projectors of the
observables $\hat{o}$ and $\hat{\omega}$ respectively. Therefore,
projection operator $\hat{P}_{\alpha}$ with $\alpha \equiv (i, a)$ takes
the form \BEQ \hat{P}_{\alpha}\equiv\hat{P}_{ia} = \hat{\pi}_i
\otimes \hat{p}_a. \EEQ
 Repeated measurements of $\hat{\Omega}$ which means
repeated simultaneous  measurements of $\hat{\omega}$ and
$\hat{o}$, determines the joint probabilities to observe
$\omega_i$ for $S$ and $o_a$ for $A$
 \BEQ\label{general}
P_{\alpha}\equiv P_{ia} = \textnormal{Tr}[\Rt(\hat{\pi}_i \otimes
\hat{p}_a)], \EEQ
 where
$\Rt$ is defined in (Eq.~\ref{stateaftetau}). In fact the numbers
$P_\alpha$ are the diagonal elements of
$\hU^{\dagger}(\hro\otimes\hR)\hU$ in the factorized basis which
diagonalizes $\hat{\omega}$ and $\hat{o}$.\\
 The whole elements of
the density matrix $\hro$ can be determined by the mapping $ \hro
\rightarrow P_{\alpha}$. Like the idea of finding a universal
observable \cite{D'Ariano}, if $\hH$ couples $S$ and $A$ properly,
this mapping will be expected to be invertible for $n\geq m$. We
shall see that even simple interactions can achieve this
condition. For given observables $\hat{\omega}$ of $S$ and
$\hat{o}$ of $A$ and for a known initial state $\hR$, the
precision of this procedure relies on the ratio between the
experimental uncertainty of $P_\alpha$ and the resulting
uncertainty on $\hro$, which can be characterized by the
determinant, $\Delta$, of the transformation (\ref{general}).
 For $\Delta=0$ it is impossible to determine $\hro$ from $P_\alpha$.
 This means, the system is unstable with
 respect to small errors made during experimental determination of
 $P_\alpha$ or equivalently $(\left<\siz\right>, \left<\sz\right>, \left<\sz\siz\right>)$.
 Therefore, the Hamiltonian $\hH$ and time interval $\tau$ should be chosen
 so as to maximize $|\Delta|$ over all possible unitary transformations.

\section{Spin-$\half$ Assistant in a Known Pure State}

In this section we illustrate the above ideas by studying a
two-level system $S$, namely, an spin-$\half$ system in a known
pure state. The density matrix of a spin-$\half$ can be
represented with the help of Pauli matrices. The determination of
$\hro$ corresponds to determination of the elements of the
polarization vector, $\vro$. we let \textrm{S} and \textrm{A}
interact during the time interval $\tau$. The observables
$\hat{\omega}$ and $\hat{o}$ to be measured are the
$\textit{z}$-components of spin of $\textrm{S}$ and $\textrm{A}$
and are determined by $\siz$ and $\sz$ respectively. The
projection operators are \BEA\label{projections} \hat{\pi}_{i} =
\half(\hat{1} + \siz),\;\;\; \hat{p}_{a} = \half(\hat{1} + \sz),
\EEA for $i$ and $a$ equal to $\pm 1$. Experiments will determine
the four joint probabilities $P_{\alpha} = {P_{++}, P_{+-},
P_{-+}, P_{--}}$. These probabilities are related to the three
real parameters $\vro$ of $\hro$ by inserting
(Eq.~\ref{projections}) and $\hR =\half(\hat{1} + \sz)$ into
(Eq.~\ref{general}). \BEQ\label{Palpha} P_{\alpha} = u_{\alpha} +
\vec{v}_{\alpha}\cdot \vro, \EEQ where
 \BEA\label{ualfa,valfa}
 u_{\alpha} = \half[\hU
(\hat{1}\otimes\hR) \hU^{\dagger}]_{\alpha,\alpha},\;\;\;
\vec{v}_{\alpha} = \half[\hU(\vhsig\otimes \hR)
\hU^{\dagger}]_{\alpha,\alpha}, \EEA with $\alpha = \{ia\} = \{++,
+-, -+, -- \}$ and matrix elements have been represented in the
standard representation of the Pauli matrices, $\vhsig$ and
$\vec{\hat{s}}$. The probabilities $P_{\alpha}$ should be positive
and normalized for any the density matrix, $\hro$, such that $\vro
\,^{2}\leq 1$. These conditions imply that
 \BEA\label{condition}
u_{\alpha}\geq \mid v_{\alpha}\mid,\;\;\;\;\;\;\;\;\sum_{\alpha} u_{\alpha} = 1,\;\;\;\;\;\;\;\; \sum_{\alpha}\vec{v}_{\alpha} =  0. \EEA
 The determinant of
transformation $\hro\rightarrow P_{\alpha}$ can be either
 \BEQ
\vec{v}_{++} \cdot (\vec{v}_{+-} \times \vec{v}_{+-}), \EEQ
 or any
other permutations of three of the vectors $\vec{v}_{++}$,
$\vec{v}_{+-}$, $\vec{v}_{-+}$ and $\vec{v}_{--}$. Therefore, the
determinant of the transformation is four times the volume of the
parallelepiped made by three of these vectors. For example, \BEQ
\Delta = 4 \vec{v}_{++} \cdot (\vec{v}_{+-} \times \vec{v}_{+-}).
\EEQ If the unitary evolution operator $\hU$ is such that vectors
$\vec{v}_\alpha$ are not coplanar, the transformation
(Eq.~\ref{Palpha}) is invertible and one can determine $\vro$ from
the set of $P_{\alpha}$. Alternatively, $\hro$ is deduced from
$\langle\siz\rangle$, $\langle\sz\rangle$ and $\langle\sz \siz\rangle$ at time $\tau$.
 The notation $\sz\siz$ is used for simplicity instead of $\sz\otimes\siz$ which
lives in the common Hilbert space of $\textrm{S}$ and
$\textrm{A}$. $\siz$, $\sz$ and $\sz\siz$ can be simultaneously
measured and are in one to one
correspondence with the set of probabilities $P_{\alpha}$.\\
We first look for the upper bound of the determinant of
transformation (Eq.~\ref{Palpha}), $|\Delta|$ implied by the
conditions (Eq.~\ref{condition}). First we note that $|\Delta|$
increases with $|\vec{v}_{\alpha}|$ for each $\alpha$. We
therefore maximize $\Delta^2$ under the constraints
\BEA\label{constraints} \sum_{\alpha}|\vec{v}_{\alpha}| =
1,\;\;\;\sum_{\alpha}\vec{v}_{\alpha}=0.\EEA
  This yields a
symmetric solution for all these vectors \BEA\label{tetrahedron}
u_{\alpha} = |\vec{v}_{\alpha}| = \frac{1}{4},\;\;\;
\cos(\vec{v}_{\alpha},\vec{v}_{\beta}) = \frac{\vec{v}_{\alpha}
\cdot \vec{v}_{\beta} }{|\vec{v}_{\alpha}| |\vec{v}_{\beta}|} = -
\frac{1}{3}. \EEA By definition this means that vectors
$\vec{v}_{\alpha}$ form a regular tetrahedron. These solutions are
not unique and they follow from one another by rotating in the
space of the spins and permutations of the indexes $\alpha$.
Therefore, the corresponding determinant for the upper bound is
 \BEQ |\Delta| = \frac{1}{12
\sqrt{3}}. \EEQ
 Having a non-zero determinant for the proposed
procedure, ensures its feasibility.\\
One simple choice for  vectors $\vec{v}_{\alpha}$ is
\BEA\label{tetarvectors}
&&\vec{v}_{++} = \frac{1}{4 \sqrt{3}} (1, 1, 1),\;\;\;\;\;\;\; \vec{v}_{+-} = \frac{1}{4 \sqrt{3}} (- 1, 1, - 1),\nonumber\\
&&\vec{v}_{-+} = \frac{1}{4 \sqrt{3}} (1, - 1, - 1),\;\;\;\;\;\;
\vec{v}_{--} = \frac{1}{4 \sqrt{3}} (- 1, - 1, 1).
 \EEA
 This yields a  simple form for the density matrix of
 \textrm{S}:
\BEA\label{densitypure}
\rho_1 =\sqrt{3}\langle\siz\rangle,\;\;\; \rho_2 = \sqrt{3}\langle\sz\rangle,\;\;\;\rho_3 = \sqrt{3}\langle\sz\siz\rangle,\EEA
 which gives directly the whole
elements of the density matrix, $\hro$, in terms of the
expectation values and the correlation of the commuting
observables $\siz$ and $\sz$ in the final state.\\
Next step is to find out the interaction Hamiltonian and
the interaction time $\tau$ which give such a description of the tested
system, $S$.\\
This correspondence can be achieved
under the action of the Hamiltonian \BEQ\label{hamiltoni}
 \hH = \frac{1}{\sqrt{2}} \six
 ( \sx \cos \phi + \sz \sin \phi )
+ \half [ (\sy - \sx ) \sin \phi + \sz \cos \phi ],\EEQ where $2
\phi$ is the angle between $\vec{v}_{++}$ and the
$\textit{z}$-axis, that is, $\cos \phi = \frac{1}{\sqrt{3}}$.
Noting that $\hH^2 = \sin^2 \chi$, where $\chi$ satisfies $\cos
\chi = 1/2 \cos \phi$, and taking as duration of the evolution
$\tau = \chi/\sin \chi$, we obtain $\hU \equiv exp(- i \hH \tau) =
\cos \chi - i \hH $. The simpler form \BEQ\label{simplehamiltoni}
\hH = \frac{1}{\sqrt{2}} \six\sx + \half \left(\sy \sin \phi +
\sz\right)\EEQ
 of $\hH$ can be obtained by a rotation
of $\vec{\hat{s}}$ and also achieves an optimal mapping
$\hro\rightarrow P_{\alpha}$, provided $\sz\rightarrow \sx
\sin\phi + \sz \cos \phi$ both in the measured projections
$\hat{p}_a = 1/2(1\pm \sz)$ and in the initial state $\hR =
\hat{p}_{+}$. The first term in (Eq.~\ref{simplehamiltoni})
describes, in the spin language, an Ising coupling, while the
second term represents a transverse magnetic field acting on the
assistant \textrm{A}.

\section{Assistant System as a Coherent State of Light}
 In this section we
discuss the possibility of using light as an assistant to
determine the  elements of the density matrix of a
spin-$\half$ system.  We show  that in case of using an
electromagnetic field in coherent state, one can determine the
state of $\textrm{S}$ from the
commutative measurements on $S$ and $\textrm{A}$.\\
To describe this physical situation we choose   the Jaynes-Cummings model
\cite{Walls, leon}
a widely accepted model describing the
interaction of matter (two-level atom or spin-$\half$ system) and
a single mode of radiation. This model  is exactly solvable but still rather
non-trivial and it finds direct experimental realization in
quantum optics. The Hamiltonian reads \cite{Walls}:
\begin{equation}\label{Hamiltonian}
\hat{H}=\hat{H}_A+\hat{H}_S+\hat{H}_{SA}=\hbar \omega \ad \add +\frac{1}{2}\hbar \omega
\hat{\sigma}_z +\hbar \gamma (\spp \add +\smm \ad)
\end{equation}
where $\ad$ and $\add$ are the standard photon creation and
annihilation operators of the field (the assistant $\textrm{A}$), with commutation
relation $[\add,\ad]=1$,  $\hsig_i$ are the standard Pauli matrices for the spin of the two level system $S$.
The total Hamiltonian (\ref{Hamiltonian}) is the sum of the Hamiltonian
of the field $\hat{H}_A$, the Hamiltonian $\hat{H}_S$ of the two level system $S$ and
 the interaction Hamiltonian $\hat{H}_{SA}$
 which can be written as $\hbar \gamma \hV$,
with $\gamma$  the coupling constant. It can be easily checked that
the interaction operator, $\hV$, and the total
number of excitations, $\hN$:
\BEA\label{N&V1} \hV = \spp \add + \smm \ad,\;\;\;\hN = \ad \add
+ \spp \smm. \EEA
 are two constants of motion.
 $\hV$ and $\hN$ also  commute with each other since $\hV^{2} = \hN$.

One can then calculate exactly the relevant observables at time
$t$ using the Heisenberg equations of motion:
 \BEA\label{Heisenberg}
&&\dot{\add}= - i \omega \add - i \gamma \smm,\nonumber\\
&&\dot{\hat{\sigma}}_{-} = - i \omega \smm + i \gamma \siz \add,\nonumber\\
&& \dot{\hat{\sigma}}_z = 2 i \gamma (\ad \smm - \spp \add). \EEA
 We will briefly outline the result.
The exact solution of the above set of equations reads:
\begin{eqnarray}\label{sigmat}
&&\add(t)=e^{i(\gamma \hat{V}-\omega)t}\left[\left(\cos \gamma\hat{K}t  -\frac{i \hV \sin \gamma\hK t}{\hK}\right)\add(0)-\frac{i \sin \gamma\hK t}{\hK} \smm(0)\right],\nonumber\\
 &&\smm(t)=e^{i(\gamma \hat{V}-\omega)t}\left[\left(\cos \gamma\hat{K}t  +\frac{i \hV \sin \gamma\hK t}{\hK}\right)\smm(0)-\frac{i \sin \gamma\hK t}{\hK}
 \add(0)\right],
\end{eqnarray}
where $\hK=\sqrt{\hN +1}=\sqrt{\hV^2 +1}$.
Note  that $\hK$ and $\hV$ commute  so that their mutual ordering is irrelevant.\\
We study the case in which the electromagnetic field is  in a coherent
state, a condition that coincides with the  common experimental
situation  of a resonant laser mode  interacting with a \sph system. Assuming initial factorization between $S$ and $A$, the density matrix of the total system $\textrm{S+A}$  at time $t=0$,   is:
\begin{equation}\label{coherentdensity}
  \hat{\rho}(t=0)=\frac{1}{2}(1+\langle \hat{\sigma}_i \rangle\hat{\sigma}_i) \otimes  \sum_{k=0}^{\infty} \sum_{m=0}^{\infty}
  \frac{e^{-|\alpha|^2}}{\sqrt{k!} \sqrt{m!}} \alpha^k {\alpha^*}^m  |k\rangle \langle m|
\end{equation}
where $|k\rangle$ denotes the ket for the photon quanta
and   $|\alpha|^2$ is the average  photon number in the coherent state.

We consider as possible triplets of commuting observables the following:
\begin{eqnarray}
\nonumber &&\hat{\sigma}_z\;\;, \;\; \ad \add\;\; ,\;\; \hat{\sigma}_z \ad\add\\
\nonumber &&\hat{\sigma}_x\;\;, \;\; \ad \add\;\; ,\;\; \hat{\sigma}_x \ad\add
\end{eqnarray}
but  others combinations are  possible.
A direct evaluation shows
 that only the 
second
 choice produces a set of independent
equations relating the measurements of the chosen commutating observables at time $t>0$, after turning on the interaction, and the state of the system $S$ at time $t=0$. Only  in this
case then, a reconstruction of the state of the \sph system at time $t=0$ is possible .  We  report here  the details of the calculation only for this
relevant choice of observables. We have to calculate the expectation values of these three
observables at time $t$. In order to perform this calculation, we will make use of the
eigenfunctions of the $\Vh$ operator, which are:
\begin{equation}
\fin{n}=\frac{|n-1\rangle |+\rangle \pm |n\rangle |-\rangle}{\sqrt{2}} \;\;\; ( n\geq1)\;,
\;\;\;\;\; \fin{0} =|\phi_0\rangle=|0\rangle |-\rangle  \;\;\;\;\;\;(n=0),\end{equation} and the operators $\Vh$ and $\Kh$, when
applied to these functions, evaluate to:
\begin{eqnarray}\label{eigenvalues}
\Vh \fin{n} =\pm \sqrt{n} \; \fin{n}, \;\;\;\;\;\;\Kh \fin{n} =\sqrt{n+1} \; \fin{n}.
\end{eqnarray}
Let us introduce the following notation:
\begin{eqnarray}
\nonumber \hat{f}(\Vh)=e^{i(\gamma \hat{V}-\omega)t}& \Rightarrow &f^{\pm}_n=\langle \finn{n} |f(\Vh)\fin{n}=e^{i( \pm \gamma \sqrt{n}-\omega)t} \rightarrow (f^{\pm}_n)^*=\bar{f}^{\pm}_n\\
\nonumber \hat{S}(\Kh)=\frac{i \sin \gamma\Kh t}{\Kh} &\Rightarrow &S_n= \langle \finn{n}
|S(\Kh)\fin{n}=\frac{i \sin \gamma \sqrt{n+1} t}{\sqrt{n+1} } \rightarrow S_n^*=\bar{S_n}=-S_n\\
 \nonumber \hat{g}(\Vh,\Kh)=\cos \gamma\hat{K}t +\Vh \hat{S}(\Kh)&\Rightarrow&
   g^{\pm}_n= \langle \finn{n}|\hat{g}(\Vh,\Kh)\fin{n}=\cos \gamma \sqrt{n+1}t  \pm  \sqrt{n}S_n  \rightarrow (g^{\pm}_n)^*=\bar{g}^{\pm}_n
\end{eqnarray}
where, for the  sake of compactness,  a bar in some cases is used instead of the asterisk to indicate the complex conjugation operation.
In this notation it holds that:
\begin{eqnarray}
 \spp(t)&=&( \spp g^{\dagger}(\Vh,\Kh)  - \ad S^{\dagger}(\Kh) )f^{\dagger}(\Vh)\hspace{3.4 cm} \\
\nonumber \ad \add (t)&=&  (\ad g(\Vh,\Kh) -\spp S^{\dagger}(\Kh))(
g^{\dagger}(\Vh,\Kh)\add-S(\Kh)\smm) \\
\nonumber \ad \add \spp(t) &=& (\ad g(\Vh,\Kh) -\spp S^{\dagger}(\Kh))(
g^{\dagger}(\Vh,\Kh)\add-S(\Kh)\smm)( \spp g^{\dagger}(\Vh,\Kh)  - \ad S^{\dagger}(\Kh) )f^{\dagger}(\Vh)
\end{eqnarray}
 For a generic observable $\hat{O}$
we define:
\begin{equation}
\nonumber \langle \hat{O}(t) \rangle=Tr[\hat{\rho}({\textstyle t=0})\hat{O}(t)]={ \textstyle \sum_{n=0}^{\infty}\sum_{i=\pm}} \langle \phi_n^i  |\hat{\rho}(t=0) \hat{O}(t)|\phi_{n}^{i}\rangle
\end{equation}
We can then proceed to evaluate  the expectation values at a generic time $t$ of the chosen triplet of commuting observables. As a first step  we have:
\begin{eqnarray}\label{sxx}
\nonumber \langle \hat{\sigma}_+(t)\rangle&=& \sum_{n=0}^{\infty} \eb
  \left[ \left( \frac{\fb^{+}_n \gb^{+}_n -\fb^{-}_n \gb^{-}_n}{2 \alpha}\sqrt{n}-\bar{S}_n\frac{n-|\alpha|^2}{\alpha} \frac{\fb^+_n-\fb^-_n}{2} \right)\lambda_1 - \right. \\
\nonumber &&\left.  \bar{S}_n \frac{\fb^+_n-\fb^-_n}{2}\frac{\as \sqrt{n}}{\ass} \lambda_2+  \left(\frac{\fb^{+}_n \gb^{+}_n
+\fb^{-}_n \gb^{-}_n}{2} -\sqrt{n} \bar{S}_n \frac{\fb^+_n-\fb^-_n}{2}\right)\lambda_2^*-\bar{S}_n \frac{\fb^+_n-\fb^-_n}{2} \alpha^* \right]
\\
\nonumber  \langle \ad \add(t)\rangle &=&
  \sum_{n=0}^{\infty} \eb \left[ \left(|g^{+}_n|^2(n-|\alpha|^2)
   -S_n  \sqrt{n} \frac{g^+_n-g^-_n}{2} + |S_n|^2\right)\lambda_1+\right. \\
&&\left.  S_n\frac{g^+_n+g^-_n}{2}( \ass \lambda_2^*-\as \lambda_2)+|\alpha|^2
|g^+_n|^2 \right] \hspace{-1 cm}\\
\nonumber  \langle \ad \add \hat{\sigma}_+(t)\rangle & =&\sum_{n=0}^{\infty} \eb
 \left[ \frac{n}{\ass} \left( |g^{+}_n|^2 ( \sqrt{n} \frac{\fb^{+}_n \gb^{+}_n
-\fb^{-}_n \gb^{+}_n}{2} -S_n  \frac{\fb^{+}_n \gb^{+}_n-\fb^{-}_n
\gb^{+}_n}{2}\frac{g^+_n-g^-_n}{2} + \right. \right. \\
\nonumber  &&  \left. \left. S_n   \frac{f^+_n-f^-_n}{2}(n -\frac{|\alpha|^2 (n+1)}{n} (1-\frac{1}{\lambda_1}))
  +  S_n \frac{\fb^{+}_n \gb^{+}_n +\fb^{-}_n \gb^{-}_n}{2}\frac{g^+_n+g^-_n}{2} \frac{|\alpha|^2}{n} (1-\frac{1}{\lambda_1})\right. \right.   \\
\nonumber &&  \left. \left. + |S_n|^2 ( \frac{\fb^{+}_n \gb^{+}_n -\fb^{-}_n \gb^{-}_n}{2 \sqrt{n}}+\frac{\fb^+_n+\fb^-_n}{2}S_n-\frac{n+1}{\sqrt{n}}
   \frac{g^-_n+g^+_n}{2} \frac{\fb^+_n-\fb^-_n}{2}  +2\sqrt{n}
\frac{g^+_n-g^-_n}{2}\cdot  \right. \right. \\
\nonumber &&   \left. \left.
\frac{\fb^+_n+\fb^-_n}{2}-\frac{|\alpha|^2}{\sqrt{n}}
   \frac{g^-_n+g^+_n}{2} \frac{\fb^+_n-\fb^-_n}{2}) \right)\lambda_1 + \left( n \frac{\fb^{+}_n \gb^{+}_n +\fb^{-}_n \gb^{-}_n}{2}|g^+_n|^2 +S_n\frac{\fb^+_n-\fb^-_n}{2}\cdot
  \right. \right. \\
\nonumber && \left.   \left.|g^+_n|^2  \sqrt{n} +2 S_n\frac{\fb^{+}_n \gb^{+}_n+\fb^{-}_n \gb^{-}_n}{2} \frac{\gb^+_n-\gb^-_n}{2}- |S_n|^2 (2  \frac{g^-_n-g^+_n}{2}\frac{\fb^+_n-\fb^-_n}{2}+\frac{n+1}{n} \cdot   \right. \right. \\
\nonumber && \left. \left.\frac{g^-_n+g^+_n}{2} \frac{\fb^+_n+\fb^-_n}{2}  + \frac{\fb^{+}_n \gb^{+}_n  +\fb^{-}_n \gb^{-}_n}{2n}-\frac{\fb^+_n-\fb^-_n}{2 \sqrt{n}}S_n)\right) \lambda_2^* +n \frac{\as}{\ass} \left( |S_n|^2 \frac{g^-_n+g^+_n}{2} \cdot  \right.  \right.\\
 \nonumber &&\left.  \left.
\frac{\fb^-_n+\fb^+_n}{2}+  S_n (|g^+_n|^2 \frac{\fb^+_n-\fb^-_n}{2}\frac{n+1}{\sqrt{n}} -\frac{\fb^{+}_n \gb^{+}_n -\fb^{-}_n \gb^{-}_n}{2} \frac{g^+_n+g^-_n}{2 \sqrt{n}})  \right) \lambda_2  \right]
    \end{eqnarray}
where we have also defined:
\begin{equation}
 \lambda_1=\frac{1+\langle \hat{\sigma}_z(0)\rangle}{2}, \hspace{0.5 cm}
\lambda_2= \langle \hat{\sigma}_+(0) \rangle,  \hspace{0.5 cm} \lambda_2^*=\langle \hat{\sigma}_-(0) \rangle
\end{equation}
From Eqs.~(\ref{sxx}) it is then easy to derive a system of three
equations relating the three expectation values:
\begin{equation}\label{system}
\langle \hat{\sigma}_x(t)\rangle =2 \Re[\langle \hat{\sigma}_+(t)\rangle],\;\;\;\;
  \langle \ad \add(t)\rangle, \;\;\;\;\langle \ad \add \hat{\sigma}_x(t)\rangle=2\Re \left[\langle \ad \add \hat{\sigma}_+t)\rangle \right]
\end{equation}
evaluated at a generic time $t$, to  the variables
$\langle\six(0)\rangle,\langle\siy(0)\rangle,\langle\siz(0)\rangle$
, i.e.
 to the density matrix of the  \sph system at time $t=0$.
In order to assess if this system of equations has solutions, we have to evaluate the determinant of the matrix $\hat{M}$  made up by the  coefficients of
$\langle\six(0)\rangle,\langle\siy(0)\rangle,\langle\siz(0)\rangle$
appearing in the system of equations determined  by the evaluation of
 (\ref{system}).
In order to achieve this goal,  it is convenient to define $\hat{M}_{ij}=\sum_{n=0}^\infty \eb  M_{ij}(n)$  and then determine
the coefficients $M_{ij}(n)$.
 Name  $A_{ij}(n)$ the same type of  coefficients  but relative to the matrix $\hat{A}$ made up by the  coefficients  multiplying the variables $\lambda_1, \lambda_2, \lambda_2^*$   in the system of equations  (\ref{sxx}).  By replacing the definitions of $f^{\pm}_n, g^{\pm}_n, S_n$ and then simplifying, one gets:
\begin{eqnarray}
A_{11}(n)&=&-i \frac{e^{i \omega t}}{\ass}(\sqrt{n}\cos{\gamma\sqrt{n+1}t}\sin{\gamma\sqrt{n}t}+|\alpha|^2\frac{\cos{\gamma\sqrt{n}t}\sin{\gamma\sqrt{n+1}t}}{\sqrt{n+1}})\\
\nonumber A_{12}(n)&=&\frac{i e^{i \omega t} \sqrt{n}}{\ass \sqrt{n+1}}\sin{\gamma\sqrt{n+1}t}\sin{\gamma\sqrt{n}t}\\
\nonumber A_{13}(n)&=&e^{i \omega t} \cos{\gamma\sqrt{n+1}t}\cos{\gamma\sqrt{n}t}\\
\nonumber A_{21}(n)&=&\frac{(1+2n)(1+n-|\alpha|^2)-(1+n+|\alpha|^2 )\cos{2 \gamma\sqrt{n+1} t}}{2 (n+1)}\\
\nonumber A_{22}(n)&=&\frac{i \ass \cos{\gamma\sqrt{n+1}t}\sin{\gamma\sqrt{n}t}}{\sqrt{n+1}},\;\;\;\;A_{23}(n)=A_{22}^*(n)\\
\nonumber A_{31}(n)&=&n A_{11}(n),\;\;\;\;A_{32}(n)=n A_{13}(n),\;\;\;\;\; A_{33}(n)=n A_{12}(n)
\end{eqnarray}
It is easy to check that these coefficients are related to the coefficients $ M_{ij}(n)$ in the following way :
\begin{equation}
M_{i1}(n)=\Re[A_{i1}(n)],\;\;\;\;M_{i2}(n)=\Re[A_{i2}(n)+A_{i3}^*(n)],\;\;\;\;M_{i3}(n)=\Im[A_{i2}(n)+A_{i3}^*(n)]
\end{equation}
If we now  calculate the determinant $\Delta(t)$ of the matrix $\hat{M}$
, we obtain:
\begin{equation}
\Delta(t)=\sum_{l=0}^{\infty} \sum_{m=0}^{\infty} \sum_{n=0}^{\infty}\ebbb  \eps_{ijk} M_{1i}(l)M_{2j}(m)M_{3k}(n)
\end{equation}
We  evaluate this determinant numerically.  Convergence within the seventh significant figure is reached by  keeping in each of the three sums the first  $30$ terms when $|\alpha|^2 \leq 9$. For larger values of $|\alpha|^2$ convergence
turns out to be much slower.
    In  Fig.~1 the temporal evolution of the determinant is shown.
We see that as  $|\alpha|^2$ increases, i.e. for larger average number of photons, the determinant has   fluctuations of larger amplitude, so that a time with a large enough determinant can be chosen to solve for the initial
state of the \sph system.
When the matrix elements are determined with some experimental uncertainty
    such a choice is a sensible one that
 allows a more  accurate determination of the initial state of the \sph system
and avoids cases of ill-conditioned matrix inversion.


\vspace{1 cm}
\begin{figure}[h]\label{fig1}
\centerline{
\includegraphics[width=8.5cm, height=6.2cm,angle=0]{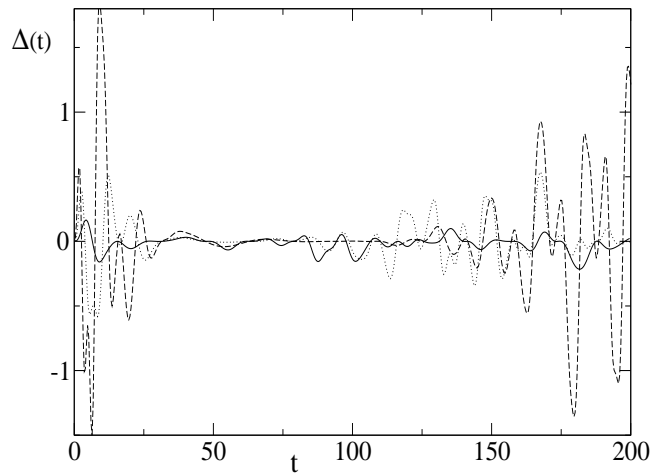}
} \caption{ Evolution of the determinant for $0<t<200$, with the
following choice  of  parameters of the model: $\gamma = 0.1,
\omega=0.1$. The continuous line  refers to the case
$|\alpha|^2=1$ the dotted line to $ |\alpha|^2=4$, the  dashed
line to $|\alpha|^2=9$, with $\alpha$ real and positive. }
\end{figure}
\section{Conclusions}
We have illustrated  a procedure which allows to reconstruct the state of a
\sph system with a simultaneous measurement of the expectation values of three commuting observables,  by coupling the system to  an assistant.
We have also illustrated how the procedure works in the simple case
of a \sph system coupled to a coherent  laser field. We have shown that in this case, after a proper choice of the commuting observables,
it is always possible to reconstruct the initial state of the \sph system. A radiation source with a large average number of photons allows to
implement the procedure in an experimentally reliable condition, i.e. with a large absolute
value of the determinant of the matrix connecting the expectation values of the commuting observables
at time $t$ to the initial state of the \sph system.
\section*{Acknowledgments}
The authors would like to thank Th. M. Nieuwenhuizen, A. E.
Allahverdyan and R. Balian for useful discussions and
proofreading. The research of G. Aquino was supported by the EC
Network DYGLAGEMEM.

\bibliographystyle{ws-procs95x65}
\bibliography{ws-pro-sample}
\end{document}